\documentclass[twocolumn,showpacs,aps,prl,superscriptaddress]{revtex4}

\usepackage{graphicx}
\usepackage{dcolumn}
\usepackage{amsmath}
\usepackage{epsfig}

\input pubboard/babarsym
\newcommand{\beq}{\begin{equation}}
\newcommand{\eeq}{\end{equation}}
\newcommand{\bea}{\begin{eqnarray}}
\newcommand{\eea}{\end{eqnarray}}

\newcommand{\Kspp}          {\mbox{$\Ks \pi^+ \pi^-$}}

\newcommand{\DE}{\ensuremath{\Delta E}}

\newcommand{\pvec}{{\bf p}}

\def\Y#1S{{\Upsilon\rm(#1S)}}
\def\Ks{{K^0_{\scriptscriptstyle S}}}

\newcommand{\BABARPubYear}    {04}
\newcommand{\BABARPubNumber}  {29}

\newcommand{\SLACPubNumber} {10650}

\def\figurebox#1#2#3{%
    \def\arg{#3}%
    \ifx\arg\empty
    {\hfill\vbox{\hsize#2\hrule\hbox to #2{\vrule\hfill\vbox to #1{\hsize#2\vfill}\vrule}\hrule}\hfill}%
    \else
    {\hfill\epsfbox{#3}\hfill}%
    \fi}

\begin{document}

\preprint{\babar-PUB-\BABARPubYear/\BABARPubNumber} 
\preprint{SLAC-PUB-\SLACPubNumber} 

\begin{flushleft}
\babar-PUB-\BABARPubYear/\BABARPubNumber\\
SLAC-PUB-\SLACPubNumber\\
\end{flushleft}

\title{
{\large \bf \boldmath
Measurements of Neutral \B\ Decay Branching Fractions \\ to \Kspp\ Final States} 
}

%
\author{B.~Aubert}
\author{R.~Barate}
\author{D.~Boutigny}
\author{F.~Couderc}
\author{J.-M.~Gaillard}
\author{A.~Hicheur}
\author{Y.~Karyotakis}
\author{J.~P.~Lees}
\author{V.~Tisserand}
\author{A.~Zghiche}
\affiliation{Laboratoire de Physique des Particules, F-74941 Annecy-le-Vieux, France }
% ...
\collaboration{The \babar\ Collaboration}
\noaffiliation

\date{\today}

\begin{abstract}
Branching fraction measurements using $B$-meson decays to $\Ks\pi^+\pi^-$ are presented. These measurements were obtained by analyzing a data sample of 88.9 million $\Upsilon(4S) \rightarrow$ \BB\ decays collected with the \babar\ detector at the SLAC PEP-II asymmetric-energy $B$ factory. Using a maximum likelihood fit, the following branching fraction results were obtained: ${\cal B}$($B^0 \to K^0\pi^+\pi^-$) = (43.7 $\pm$ 3.8 $\pm$ 3.4) $\times$ 10$^{-6}$, ${\cal B}$($B^0$ $\rightarrow$ $K^{*+}$$\pi^-$) = (12.9 $\pm$ 2.4 $\pm$ 1.4) $\times$ 10$^{-6}$ and ${\cal B}$($B^0$ $\rightarrow$ $D^-(\rightarrow \Ks\pi^-)\pi^+$ = (42.7 $\pm$ 2.1 $\pm$ 2.2) $\times$ 10$^{-6}$. The $CP$ violating charge asymmetry ${\cal A}_{K^*\pi}$ for the decay $B^0$ $\rightarrow$ $K^{*+}$$\pi^-$ was measured to be ${\cal A}_{K^*\pi} = 0.23 \pm 0.18^{+0.09}_{-0.06}$. For all these measurements the first error is statistical and the second is systematic.
\end{abstract}

\pacs{13.25.Hw, 12.15.Hh, 11.30.Er}

\maketitle

Three-body decays of the $B$ meson tend to be dominated by intermediate quasi two-body charmed particles with the charmless resonant and non-resonant contribution being small. Nevertheless, these charmless decays prove to be important in furthering our understanding of the weak interaction and complex quark couplings described by the Cabibbo-Kobayashi-Maskawa matrix elements~\cite{ckm}.

The $B$-meson decay to $\Ks\pi^+\pi^-$ can proceed via many interesting charmless resonances which we can probe for $CP$ violation, such as $f_0\Ks$~\cite{f0Ks}, $\rho^0\Ks$ and $K^{*+}\pi^-$. A limit on the sum of their branching fractions can be obtained by measuring the inclusive charmless branching fraction of $B^0 \to \Ks\pi^+\pi^-$. This measurement has been performed previously by the CLEO~\cite{cleo} and Belle~\cite{belle} experiments. For the mode $B^0 \to K^{*+}\pi^-$ the branching fraction can be measured directly with the available \babar\ data sample.

Branching fraction and asymmetry measurements of charmless $B$ decays can also be used to test the accuracy of QCD factorization models~\cite{beneke}. In particular there are factorization models that predict $CP$ asymmetries in the decay $B^0 \rightarrow K^{*+}\pi^-$~\cite{guo}. The decay $B^0 \rightarrow K^{*+}\pi^-$ is self tagged (the charge of the kaon reflects the flavor of the $B$-meson), so the $CP$ asymmetry can be defined as:

\begin{equation}
{\cal A}_{K^*\pi} = \frac{\Gamma_{K^{*-}\pi^+} - \Gamma_{K^{*+}\pi^-}}{\Gamma_{K^{*-}\pi^+} + \Gamma_{K^{*+}\pi^-}} ~.
\end{equation}

In this paper the branching fractions of $B^0 \to K^0\pi^+\pi^-$, $B^0$ $\rightarrow$ $K^{*+}$$\pi^-$ and $B^0$ $\rightarrow$ $D^-(\rightarrow \Ks\pi^-)\pi^+$ are presented, where charge conjugate decays are also implied. The procedure used selection criteria requiring events with a reconstructed $\Ks\pi^+\pi^-$ final state. In the case of  ${\cal B}(B^0 \to K^0\pi^+\pi^-$), the total charmless contribution to the Dalitz plot was measured (with charmed and charmonium resonances removed), including contributions from resonant charmless sub-structure. For the decays $B^0$ $\rightarrow$ $K^{*+}$$\pi^-$ and $B^0$ $\rightarrow$ $D^-(\rightarrow \Ks\pi^-)\pi^+$, the analysis was restricted to the region of the $\Ks\pi^+\pi^-$ Dalitz plot consistent with  $K^{*+}(\rightarrow \Ks\pi^+)$ and $D^-(\rightarrow \Ks\pi^-)$ decays respectively. Finally, the ${\cal A}_{K^*\pi}$ value for the decay $B^0$ $\rightarrow$ $K^{*+}$$\pi^-$, which was first measured by CLEO~\cite{cleoacp} was extracted.

The data used in this analysis were collected at the \pep2\ asymmetric-energy \epem\ storage ring with the \babar\ detector~\cite{babar}. The \babar\ detector consists of a double-sided five-layer silicon tracker, a 40-layer drift chamber, a Cherenkov detector, an electromagnetic calorimeter and a magnet with instrumented flux return. The data sample has an integrated luminosity of 81.9~fb$^{-1}$ collected at the $\FourS$ resonance, which corresponds to $(88.9\pm 1.0)\times 10^6$ \BB\ pairs. It was assumed that the $\FourS$ decays equally to neutral and charged $B$-meson pairs. In addition, 9.6 fb$^{-1}$ of data collected at 40~MeV below the $\FourS$ resonance were used for background studies.

Candidate $B$-mesons were reconstructed from two tracks and a $\Ks$, where the $\Ks$ was reconstructed from $\pi^+\pi^-$ candidates.  Each of the two tracks that were not generated by the $\Ks$ were required to have at least 12 hits in the drift chamber, a transverse momentum greater than 100~\mevc and to be consistent with originating from the beam-spot.  These tracks were selected as pions using energy loss (\dedx) measured in the tracking system, the number of photons measured by the Cherenkov detector and their corresponding Cherenkov angle. Furthermore, the tracks were also required to fail the electron selection based on \dedx information, the ratio of energy in the calorimeter to momentum in the drift chamber and the shape of the signal in the calorimeter. The prerequisites imposed on $\Ks$ candidates were for the reconstructed mass to be within 15~MeV/c$^2$ of the nominal $K^0$ mass~\cite{pdg}, a decay vertex separated from the $B^0$ decay vertex by at least five standard deviations and a cosine of the angle between the line joining the $B$ and $\Ks$ decay vertices and the $\Ks$ momentum to be greater than 0.999. 

To characterize signal events, two kinematic and one event shape variable were used. The first kinematic variable $\DeltaE$, is the difference between the center-of-mass (CM) energy of the $B$-candidate and $\sqrt{s}/2$, where $\sqrt{s}$ is the total CM energy. The second is the beam-energy-substituted mass $\mes = \sqrt{(s/2 + \pvec_i \cdot \pvec_B)^2/E_i^2 - \pvec^2_B}$, where  $\pvec_B$ is the $B$ momentum and  ($E_i, \pvec_i$) is the four-momentum of the $\FourS$ in the laboratory frame. Using these two kinematic variables, candidates had to be in the range  $|\Delta E| <0.1 \gev$ and $5.22<\mes<5.29 \gevcc$. The event shape variable is a Fisher discriminant ($\mathcal{F}$)~\cite{Fisher}. The $\mathcal{F}$ variable was constructed from a linear combination of the cosine of the angle between the $B$-candidate momentum and the beam axis, the cosine of the angle between the $B$-candidate thrust axis and the beam axis, and the energy flow of the rest of the event into each of nine contiguous, concentric, $10^{\circ}$ cones around the
thrust axis of the reconstructed $B$~\cite{CLEOCones}.

Continuum quark production ($e^+e^-$ $\rightarrow$ $q\bar{q}$ where $q$ = {\em u,d,s,c}) was by far the dominant source of background. This was suppressed using another event-shape variable which was the cosine of the angle $\theta_T$ between the thrust axis 
of the selected $B$-candidate and the thrust axis of the rest of the event. For continuum background, the distribution of $|\cos\theta_T|$ is strongly 
peaked towards unity whereas the distribution is flat for signal events. Therefore, the relative amount of continuum background was reduced by requiring that all candidates fulfill the criterion $|\cos\theta_T| < 0.9$.

Simulated Monte Carlo (MC) events were used to study background from other $B$-meson decays. The largest potential $B$-background was seen to come from quasi two-body decays including charmonium mesons such as $J/\psi\Ks$, $\chi_{c0}\Ks$ and $\psi(2S)\Ks$ where the charmonium meson decays to $\mu^+\mu^-$ which are misidentified as pions or where they decay directly to $\pi^+\pi^-$. These background events were removed by vetoing reconstructed $\pi^+\pi^-$ masses consistent with 3.04 $<$ $m_{\pi^+\pi^-}$ $<$ 3.17~GeV/$c^2$, 3.32 $<$ $m_{\pi^+\pi^-}$ $<$ 3.53~GeV/$c^2$ and 3.60 $<$ $m_{\pi^+\pi^-}$ $<$ 3.78~GeV/$c^2$, identifying the $J/\psi$, $\chi_{c0}$ and  $\psi(2S)$ mesons respectively. Additionally, in order to measure the charmless branching fraction of the decay $B^0 \to K^0\pi^+\pi^-$, $B^0$ $\rightarrow$ $D^-(\rightarrow \Ks\pi^-)$$\pi^+$  events were removed by vetoing events with a  reconstructed $\Ks\pi$ invariant mass consistent with  1.83 $<$ $m_{\Ks\pi}$ $<$ 1.90~GeV/$c^2$. Monte Carlo simulation showed that 21 $\pm$ 3 $B^0$ $\rightarrow$ $D^-(\rightarrow \Ks\pi^-)$$\pi^+$ background events still remained. These events had a reconstructed $D^-$ mass outside the veto as a result of using the wrong $\Ks$ or $\pi^+$ which was incorrectly selected from the other $B$ decay in the event. When selecting $B^0$ $\rightarrow$ $K^{*+}$$\pi^-$ or $B^0$ $\rightarrow$ $D^-(\rightarrow \Ks\pi^-)\pi^+$  candidates, the additional cuts 0.79 $<$ $m_{\Ks\pi}$ $<$ 0.99~GeV/$c^2$ and 1.85 $<$ $m_{\Ks\pi}$ $<$ 1.89~GeV/$c^2$ were applied respectively to the reconstructed $m_{\Ks\pi}$ invariant mass. After the above selection criteria were applied, a small proportion of events for all decays under study had more than one candidate which satisfied the selection criteria. For these events, one candidate alone was selected by choosing the candidate whose  $\cos\theta_T$ value was closest to 0. In a signal MC study, this selects the true signal candidate more than 75\% of the time.

After all cuts, the largest remaining $B$-background to $B^0 \to K^0\pi^+\pi^-$  was the 4-body decay $B^0$ $\rightarrow$ $\eta^\prime\Ks$ with $\eta^\prime$ $\rightarrow$ $\rho^0(770)\gamma$ and $\rho^0$ $\rightarrow$ $\pi^+\pi^-$ which contributes 22 $\pm$ 6 events. For the $B^0$ $\rightarrow$ $K^{*+}$$\pi^-$ and $B^0$ $\rightarrow$ $D^-(\rightarrow \Ks\pi^-)\pi^+$ channels, the background contribution was small and came from modes that can interfere by decaying to a $\Ks\pi^+\pi^-$ final state such as $f_0\Ks$ and $\rho^0\Ks$. In addition the $K^{*+}$$\pi^-$ and $D^+$$\pi^-$ modes are backgrounds to each other. Furthermore, there was the non-resonant $\Ks\pi^+\pi^-$ background contribution to the resonant signal. Along with selection efficiencies obtained from MC, using available information on exclusive measurements~\cite{hfag} or by fitting to regions in the Dalitz plot, upper limits or branching fractions for these modes were obtained to estimate their background contributions.

\begin{figure}[htb]
\resizebox{\columnwidth}{!}{
\begin{tabular}{cc}
\includegraphics{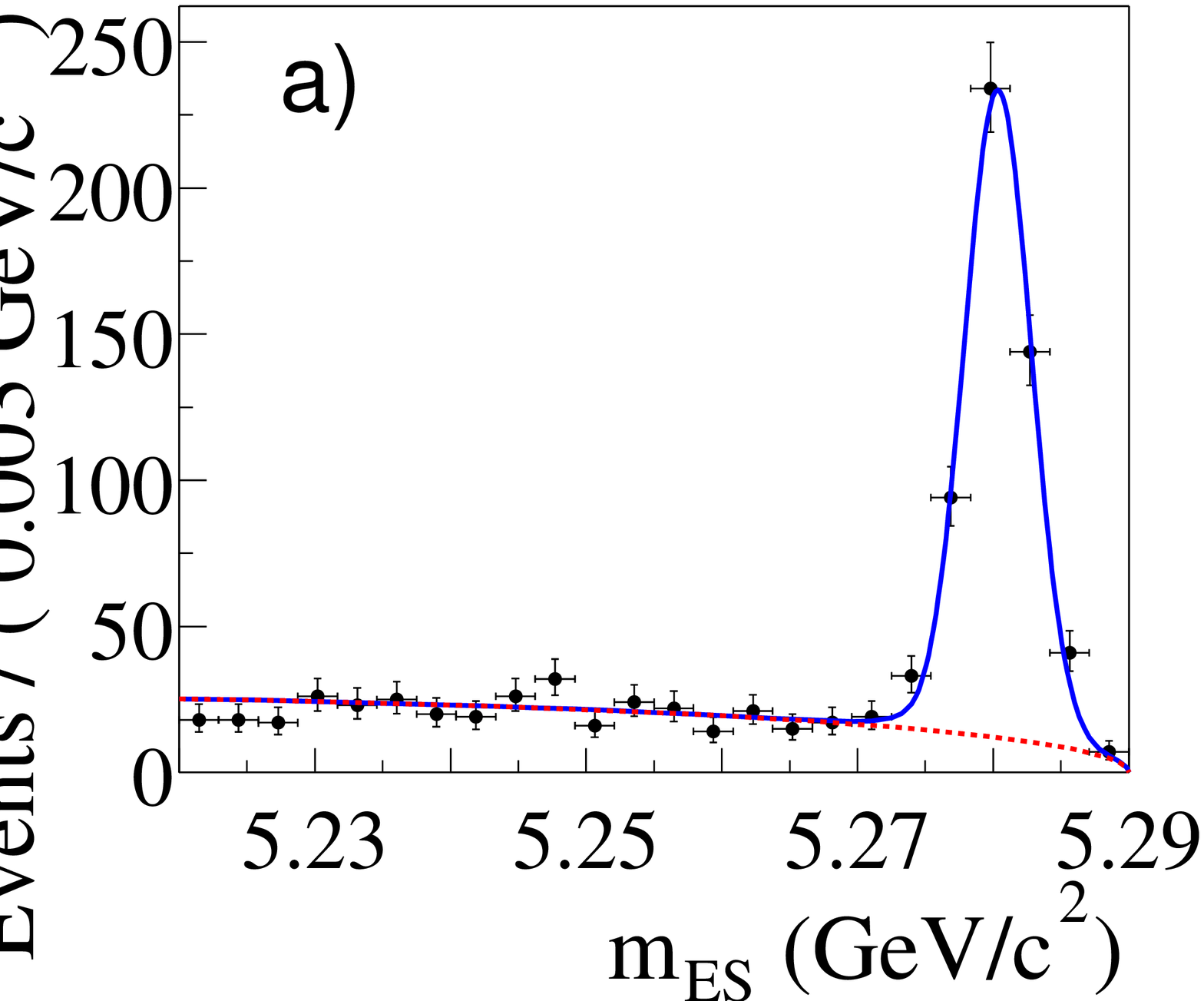} &
\includegraphics{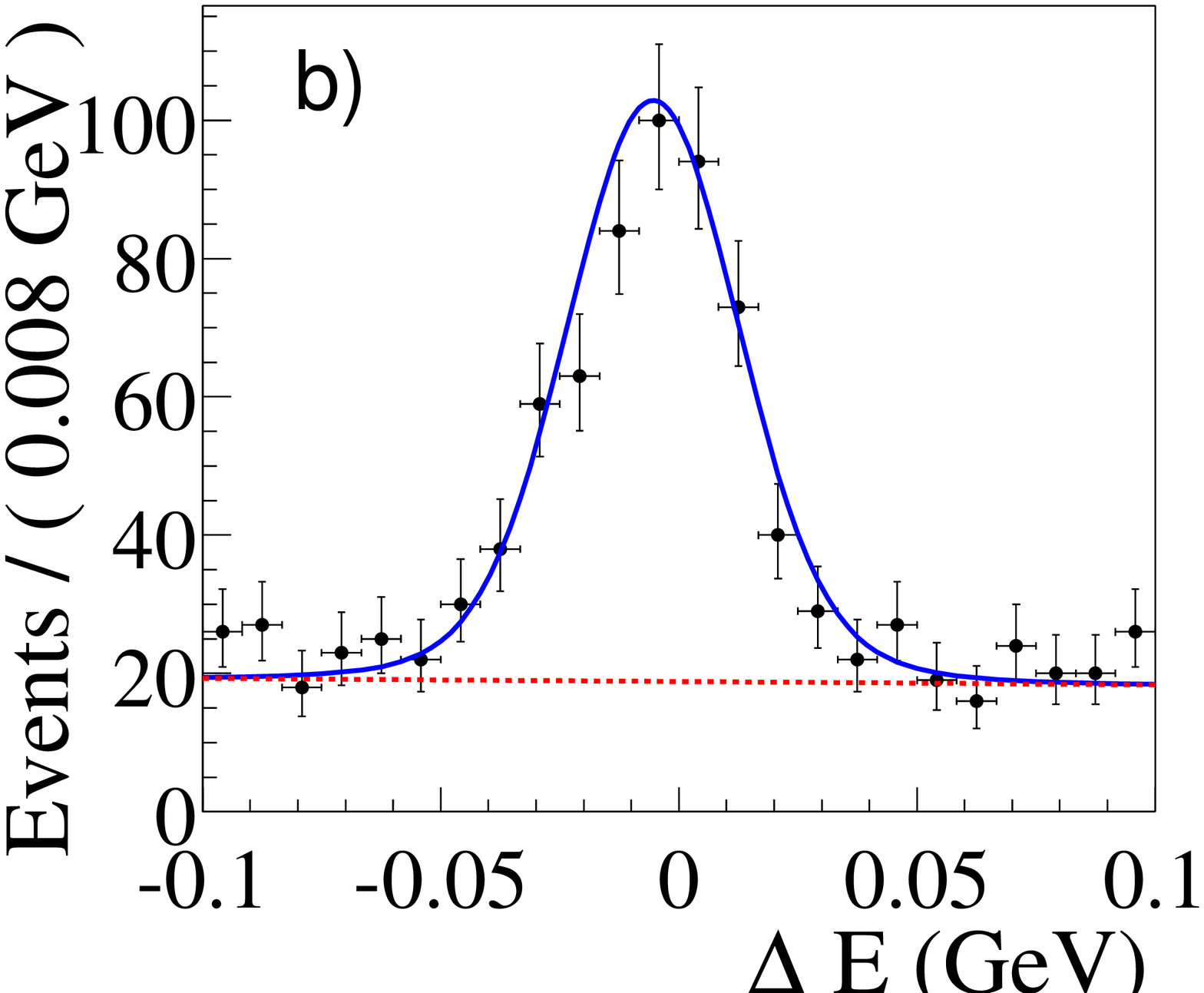}\\
\multicolumn{2}{c}{\includegraphics{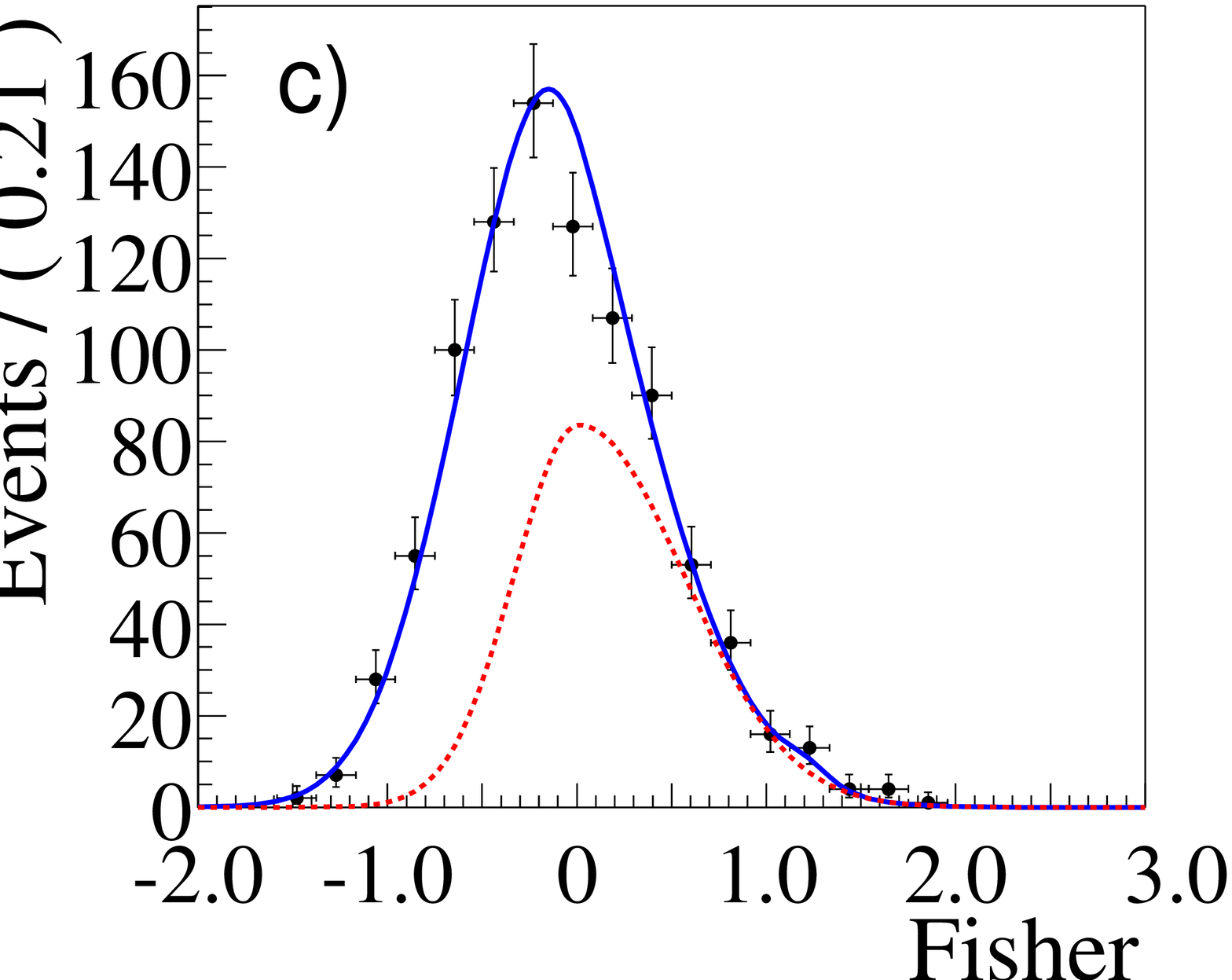}}
\end{tabular}
}
\caption{Maximum likelihood fit projections of  $m_{ES}$ (a), \DE\ (b) and  $\cal{F}$ (c) to the full set of $B^0$ $\rightarrow$ $D^-(\rightarrow \Ks\pi^-)\pi^+$ candidates. The dashed line is the fitted background PDF while the solid continuous line is the sum of the signal and background PDFs. The solid dots are data used in the fit.}\label{fig:dpi}
\end{figure}

In order to extract the signal event yield for the channel under study, an unbinned extended maximum likelihood fit was used. The likelihood function for $N$ candidates is:

\begin{equation}
  \label{eq:Likelihood}
  \mbox{$\mathcal{L}$} \,=\, \exp\left(-\sum_{i=1}^{M} n_i\right)\, \prod_{j=1}^N 
\,\left(\sum_{l=1}^M n_{l} \, P_{l}(\vec{\alpha},\vec{x_j})\right) ~,
\end{equation}

\noindent where $i$, $j$ and $l$ are integers, $M$ is the number of hypotheses (signal, continuum background and $B$-background), $P_{l}(\vec{\alpha},\vec{x_j})$ is a probability density function (PDF) with the parameters $\vec{\alpha}$ depending on three variables ($\vec{x}$) \mes, \DE, and $\mathcal{F}$, and $n_l$ is the number of events for each hypothesis determined by maximizing the likelihood function. The PDF is a product $P_{l}(\vec{\alpha},\vec{x_j}) = P_{l}(\alpha_{\mes} ,\mes ) \cdot P_{l}(\alpha_{\DE} ,\DE ) \cdot  P_{l}(\alpha_{\mathcal{F}}, \mathcal{F})$. Correlations between these variables were small for signal and continuum background hypotheses. However for $B$-background, correlations were observed between \mes and \DE, which were taken into account by forming a 2-dimensional PDF for these variables. The parameters of the signal and $B$-background PDFs were determined from MC. The continuum background parameters were allowed to vary in the fit, to help reduce systematic effects from this dominant event type. Upper sideband data defined to be in the region $0.1 < \Delta E <0.3 \gev$ and $5.22<\mes<5.29 \gevcc$ was used to model the continuum background PDFs. For the \mes\ PDFs a Gaussian distribution was used for signal and a threshold function~\cite{argus} was used for continuum. For the \DE\ PDFs a sum of two Gaussian distributions with the same means was used for the signal and a first order polynomial was used for the continuum background. Finally, for the $\cal{F}$ PDFs, a sum of two Gaussian distributions with distinct means and widths was used for signal and an asymmetric Gaussian which has different widths above and below the modal value was used for continuum background. In the case of $B$-background parameterizations, signal-like or continuum-like PDFs were used depending on the characteristics of the background. With more than 400 signal events and typically a one-to-one signal to background ratio in the total number of $B^0$ $\rightarrow$ $D^-(\rightarrow \Ks\pi^-)\pi^+$ candidates, it was possible also to vary the signal PDF parameters in the fit for this mode. This enabled uncertainties and corrections due to MC, to be calculated and applied to the  $B^0 \to K^0\pi^+\pi^-$ and  $B^0$ $\rightarrow$ $K^{*+}$$\pi^-$ analyses. Figure~\ref{fig:dpi} shows the fitted projections of the maximum likelihood fit to  $B^0$ $\rightarrow$ $D^-(\rightarrow \Ks\pi^-)\pi^+$ candidates in $m_{ES}$, \DE\ and  $\cal{F}$ containing 472 $\pm$ 24 signal and 455 $\pm$ 23 background candidates.

To extract the branching fractions for the decay modes $B^0$ $\rightarrow$ $K^{*+}$$\pi^-$ and $B^0$ $\rightarrow$ $D^-(\rightarrow \Ks\pi^-)\pi^+$ the following equation was used:

\begin{equation}
\label{eq:bf}
 \mbox{$\mathcal{B}$} \,=\ \frac{n_{sig}}{N_{BB} \times \epsilon}~,
\end{equation}

\noindent where $n_{sig}$ is the number of signal events fitted, $\epsilon$ is the signal efficiency obtained from MC and $N_{BB}$ is the total number of \BB\ events. For the charmless inclusive  $B^0 \to K^0\pi^+\pi^-$ branching fraction, the efficiency varies over the Dalitz plane and the distribution of events across it is {\em a priori} unknown, consequently the total efficiency is unknown. Therefore, to calculate the branching fraction, a weight was assigned to each event such that, for the $j$th event ${\cal W}_j$ = $\sum_iV_{sig,i}P_{i}(\vec{\alpha},\vec{x_j})/\sum_kn_{k}{P_{k}(\vec{\alpha},\vec{x_j})}$ where $V_{sig,i}$ is the signal row of the covariance matrix obtained from the fit~\cite{splot}. This procedure is effectively a background subtraction where these weights have the property $\sum_j{\cal W}_j = n_{sig}$. The branching fraction is then calculated as ${\cal B} = \sum_{j}{\cal W}_j/(\epsilon_{j} \times N_{BB})$  where $\epsilon_j$ is the efficiency which varies across the Dalitz plot and is simulated in small bins using high statistics MC.

\begin{figure}[htb]
\resizebox{\columnwidth}{!}{
\begin{tabular}{cc}
\includegraphics{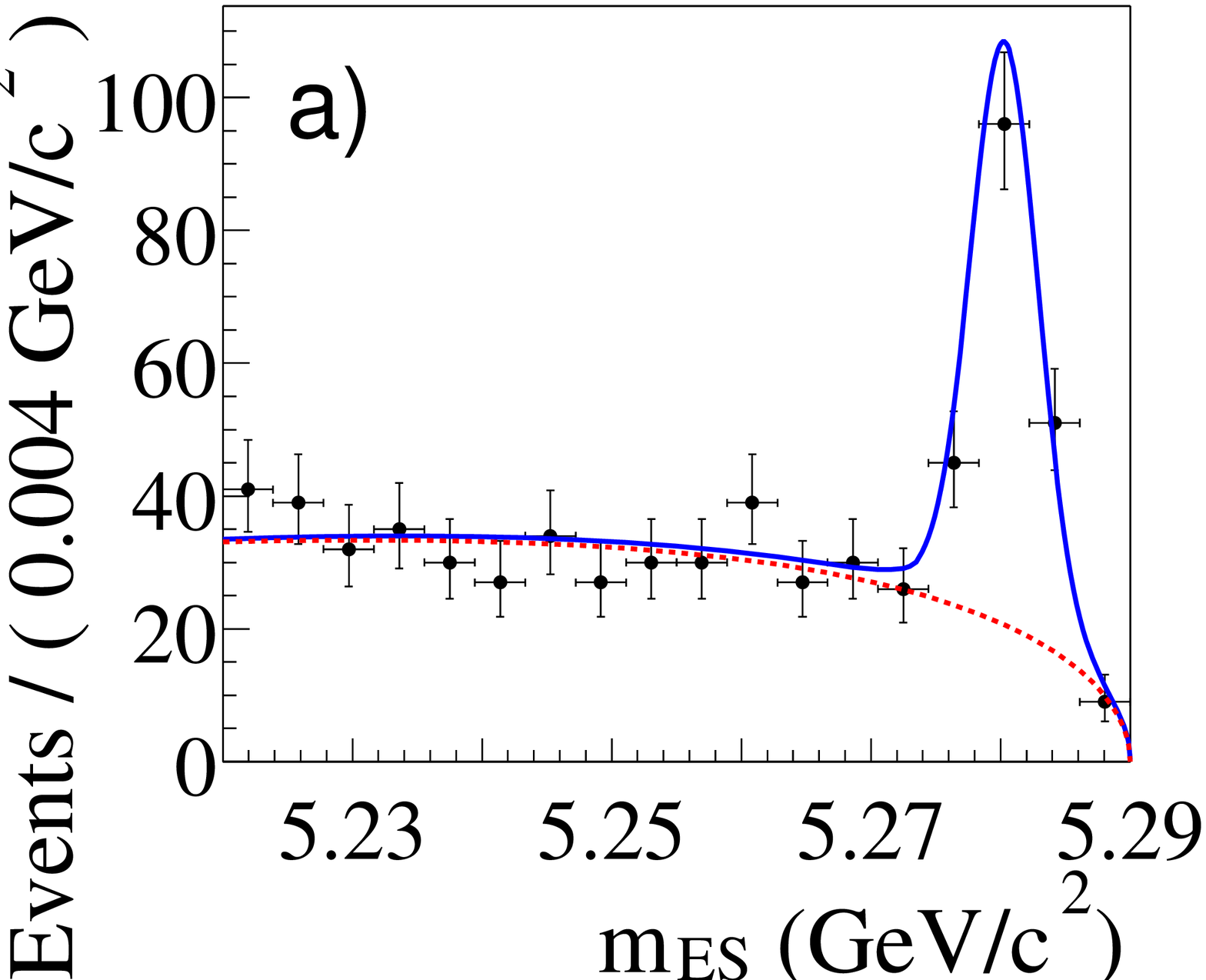} &
\includegraphics{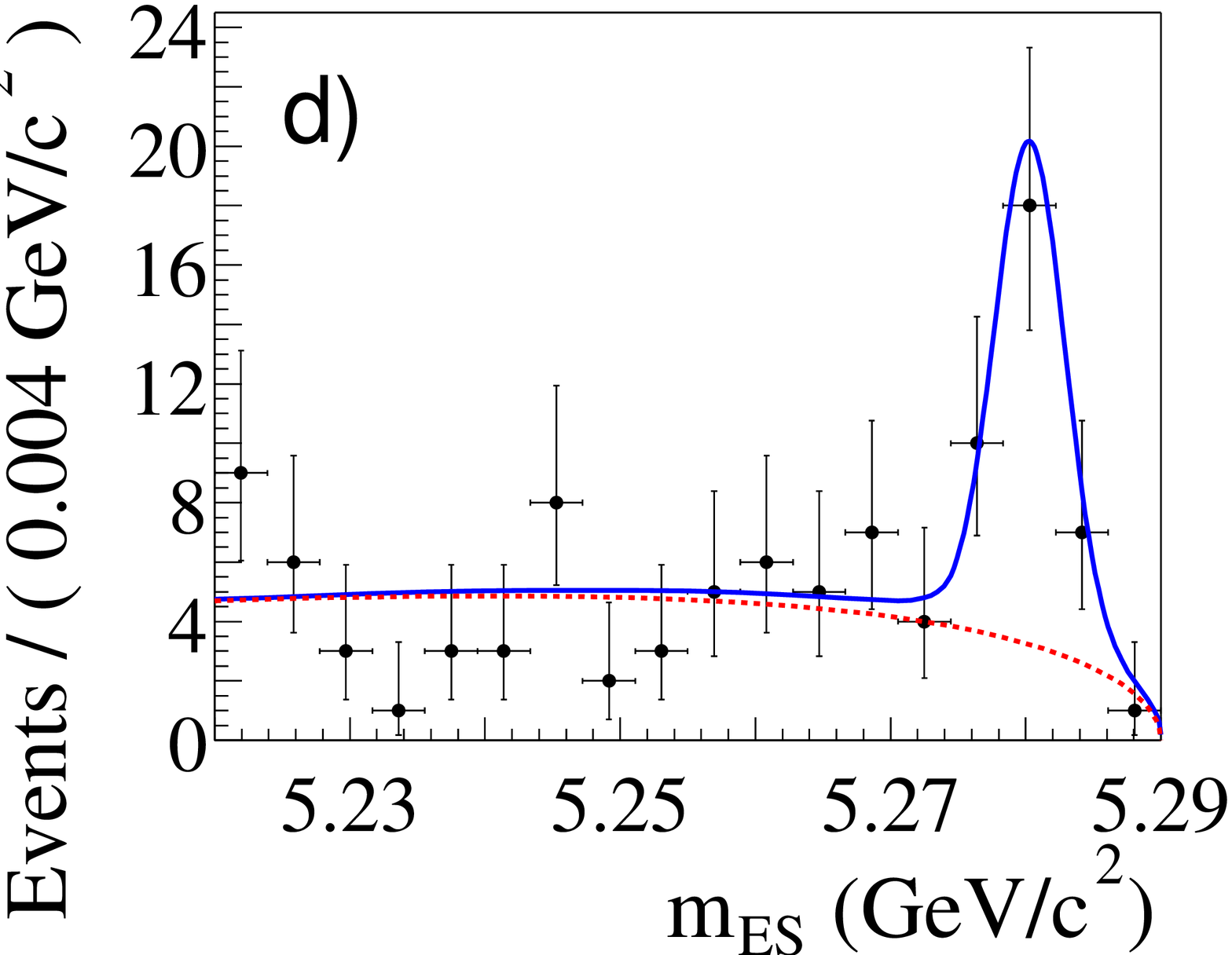}\\
\includegraphics{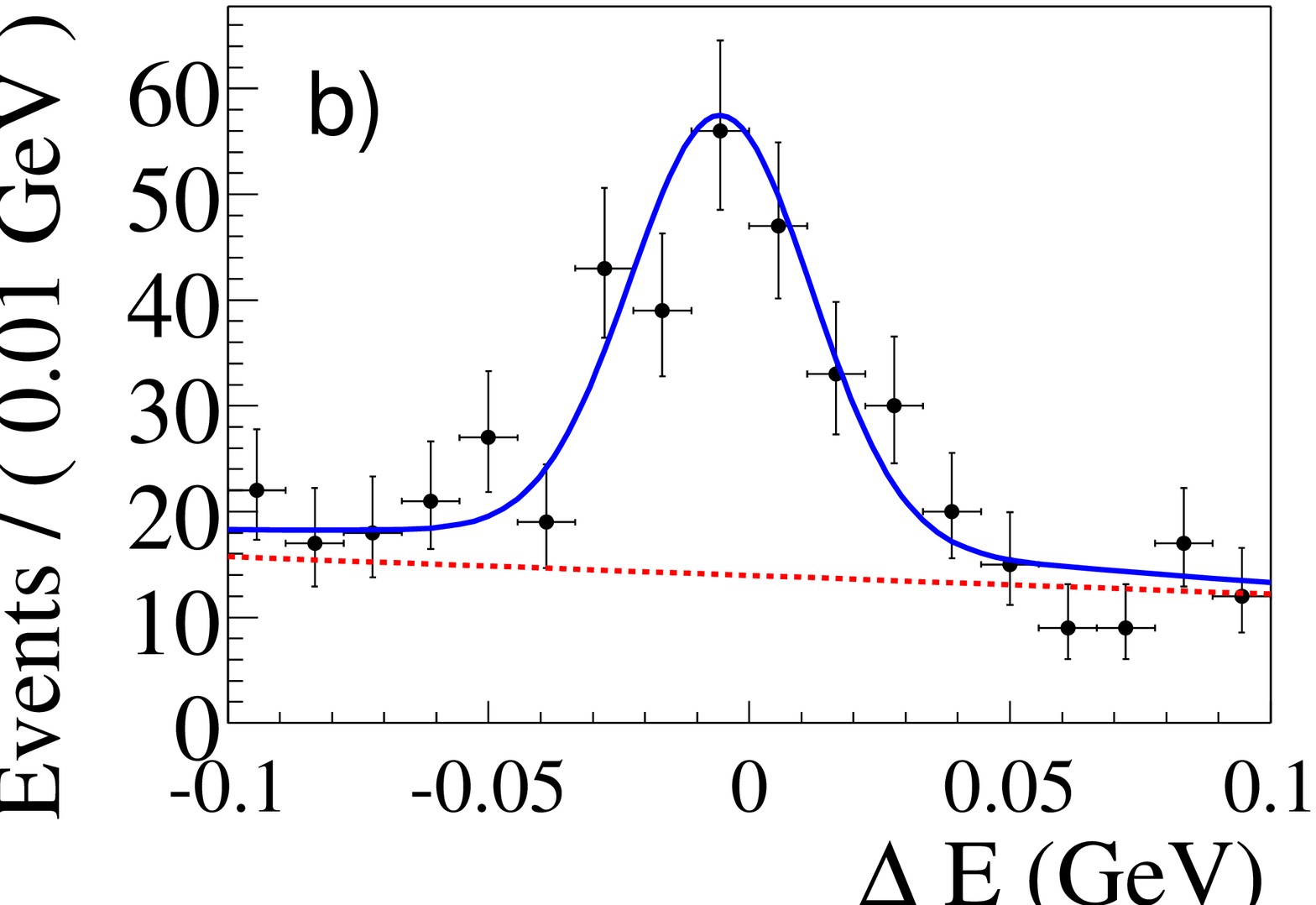} &
\includegraphics{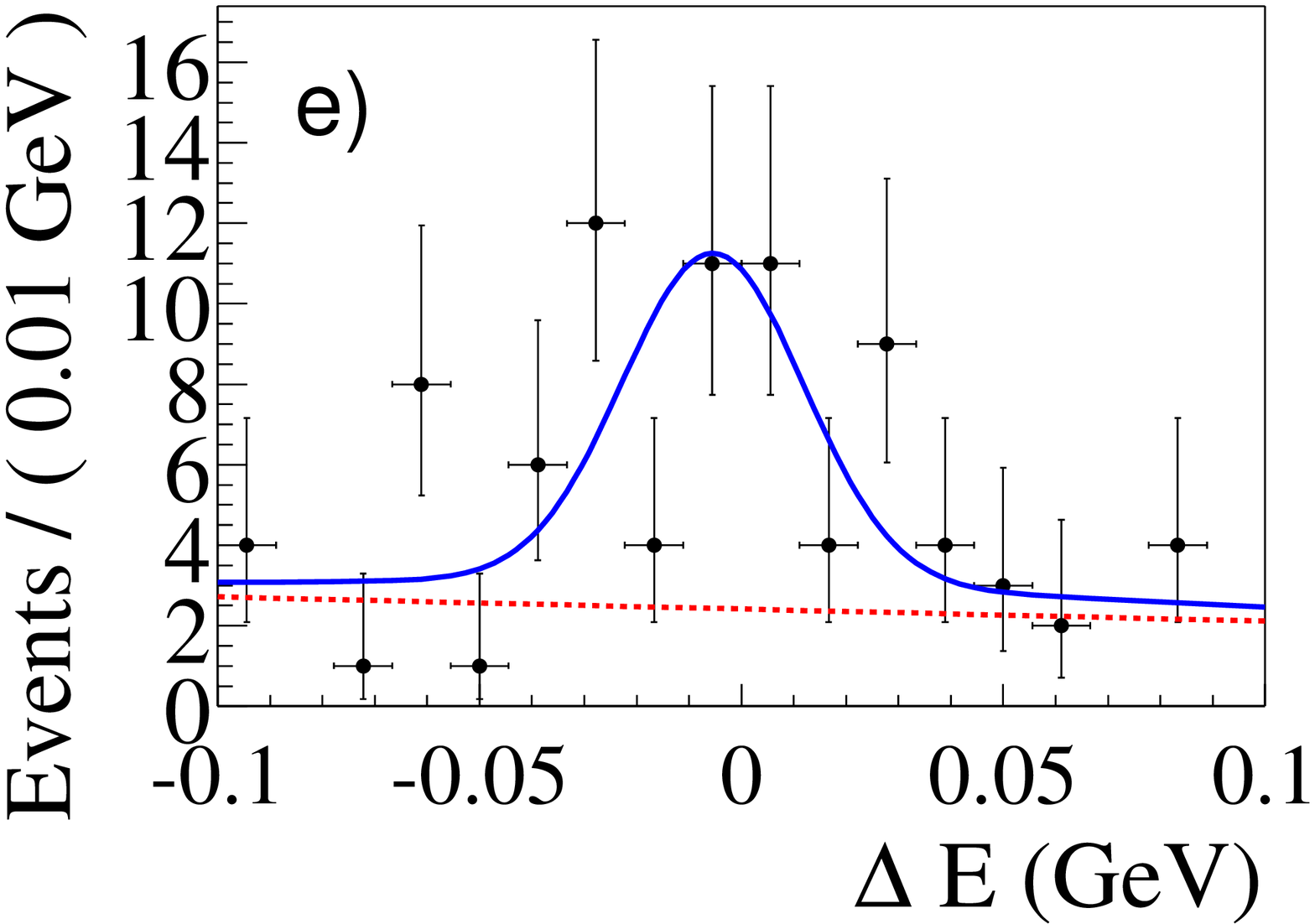}\\
\includegraphics{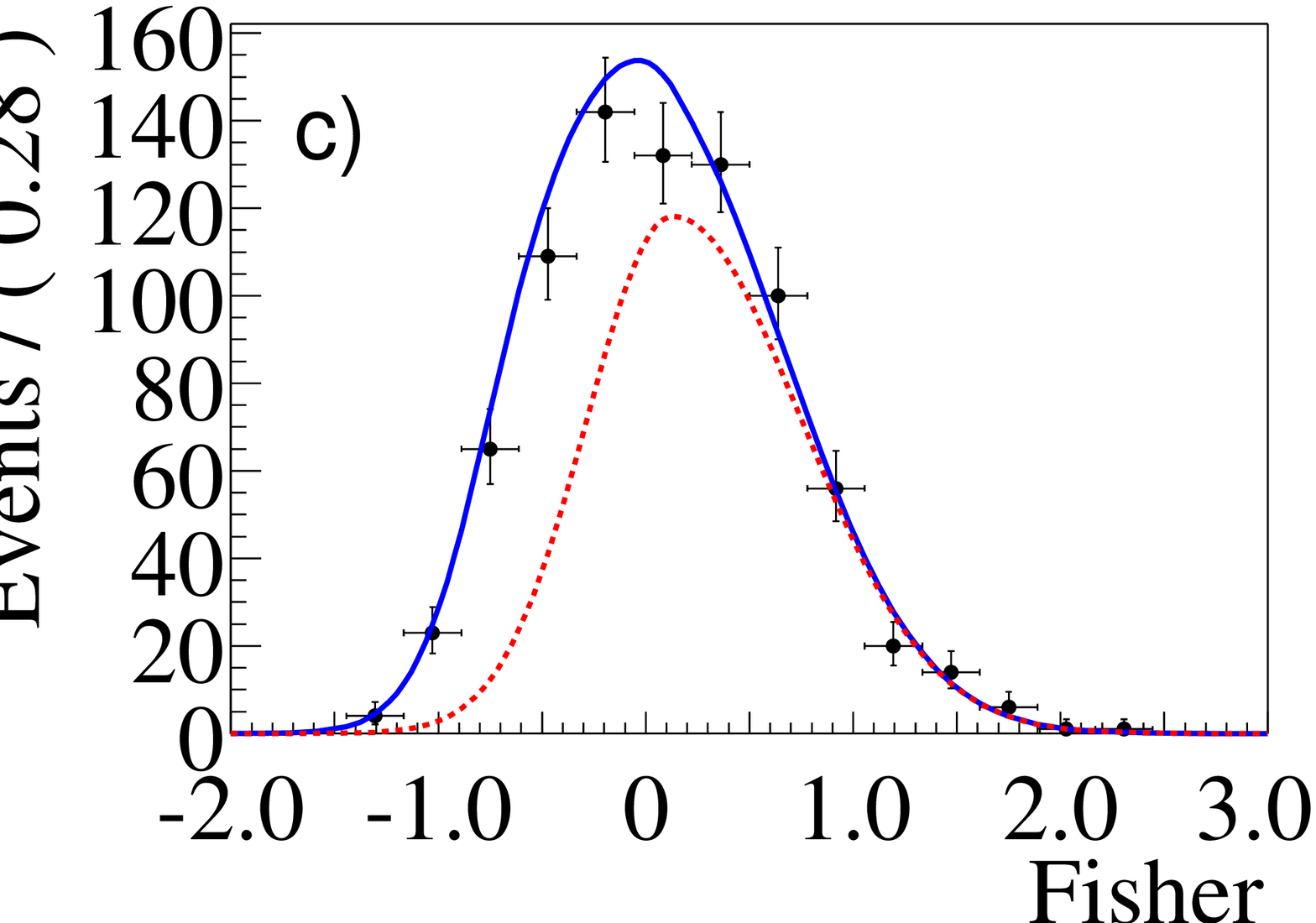} &
\includegraphics{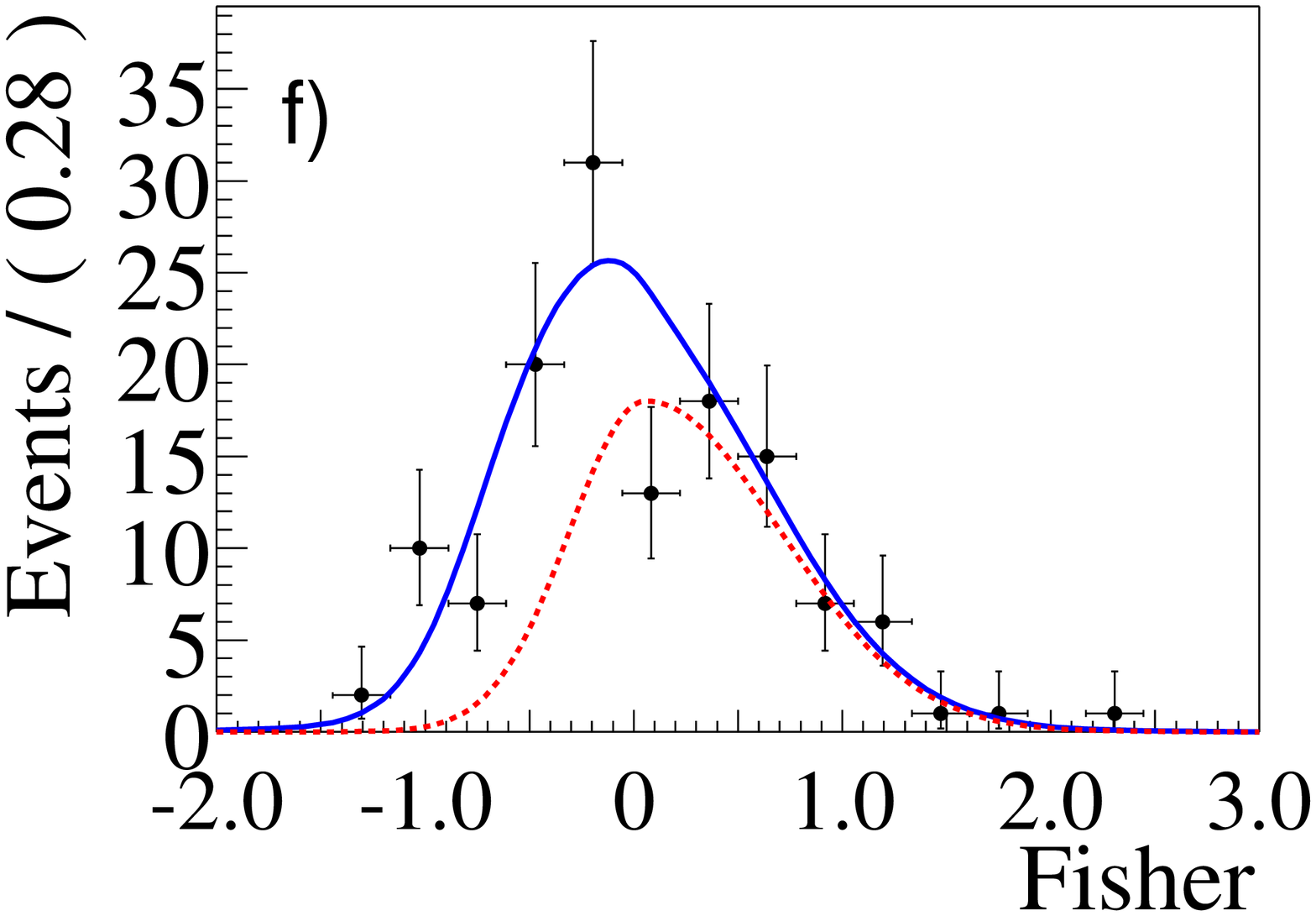}
\end{tabular}
}
\caption{Maximum likelihood fit projections of  $m_{ES}$, \DE\ and  $\cal{F}$ for signal enhanced samples of charmless $B^0 \to K^0\pi^+\pi^-$ and $B^0$ $\rightarrow$ $K^{*+}$$\pi^-$ candidates. The dashed line is the fitted background PDF while the solid continuous line is the sum of the signal and background PDFs. The solid dots are data. The left column (plots a-c) has the $B^0 \to K^0\pi^+\pi^-$ projections and the right column (plots d-f) has the $B^0$ $\rightarrow$ $K^{*+}$$\pi^-$ projections, the top, middle and bottom rows being the $m_{ES}$, \DE\, and $\cal{F}$ distributions respectively.
}\label{fig:fitproj}
\end{figure}

\begin{figure}[!htbp]
\resizebox{\columnwidth}{!}{
\begin{tabular}{c}
   \includegraphics{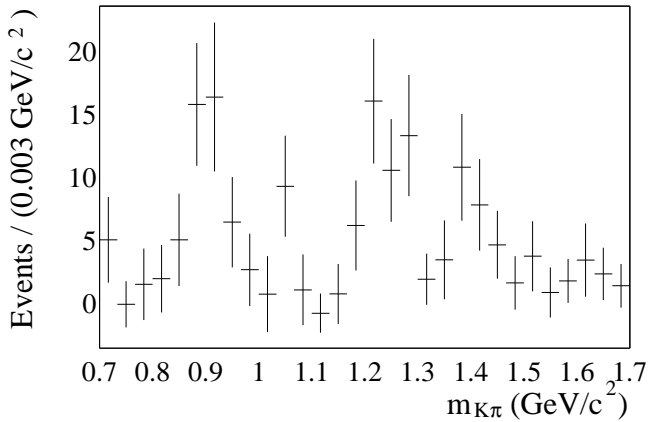} 
\end{tabular}
}
\caption{
The $m_{K\pi}$ distribution of $B^0 \to K^0\pi^+\pi^-$ candidates, weighted by $\cal{W}$ such that background events are subtracted. The one-dimensional distribution is obtained by considering events on a Dalitz plane with axes $m^2_{\Ks\pi+}$ and $m^2_{\Ks\pi-}$. The two axes are merged into one ($m^2_{K\pi}$) by folding the Dalitz plane along the line corresponding to $m^2_{\Ks\pi+}$ = $m^2_{\Ks\pi-}$ in order to obtain the above $m_{K\pi}$ mass distribution.
  }
\label{fig:projplots}  
\end{figure}

Figure~\ref{fig:fitproj} shows the fitted projections for both $B^0 \to K^0\pi^+\pi^-$ and $B^0$ $\rightarrow$ $K^{*+}$$\pi^-$ candidates, whilst the fitted signal yield and measured branching fraction are shown in Table \ref{tab:results} for all the modes under study. The average efficiency for $B^0 \to K^0\pi^+\pi^-$ signal events was approximately 8\%. Figure~\ref{fig:projplots} shows the signal mass projections of $m_{\Ks\pi}$ using $B^0 \to K^0\pi^+\pi^-$ candidates. The $m_{\Ks\pi}$ distribution clearly shows a peak at 0.9~GeV/$c^2$ which corresponds to the $K^{*+}$(892) and there is a broad structure above 1~GeV/$c^2$ which is the region where higher kaon resonances can occur.

\begin{table}[!ht]
\caption{Signal yields and branching fractions for $B^0 \to K^0\pi^+\pi^-$, $B^0$ $\rightarrow$ $K^{*+}\pi^-$ and $B^0$ $\rightarrow$ $D^-(\rightarrow \Ks\pi^-)\pi^+$ where the first error is statistical and in the case of the branching fraction measurements the second error is systematic. The efficiency of selecting $B^0$ $\rightarrow$ $K^{*+}$$\pi^-$ and $B^0$ $\rightarrow$ $D^-(\rightarrow \Ks\pi^-)\pi^+$ events was found to be 5.1\% and 12.4\% respectively. The $B^0$ $\rightarrow$ $K^{*+}\pi^-$ branching fraction takes into account that ${\cal B}(K^{*+} \to K^0\pi^+) = 2/3$, assuming isospin symmetry.}\label{tab:results}
\begin{center}
\begin{tabular}{|c|c|c|}
\hline
Mode&
Signal Events&
Branching Fraction\\
&
Yield&
($\times$ 10$^{-6}$)\\
\hline
$B^0 \to K^0\pi^+\pi^-$ &
310 $\pm$ 27&
43.7 $\pm$ 3.8 $\pm$ 3.4\\

$B^0$ $\rightarrow$ $K^{*+}\pi^-$ &
59 $\pm$ 11&
12.9 $\pm$ 2.4 $\pm$ 1.4 \\

$B^0$ $\rightarrow$ $D^-(\rightarrow \Ks\pi^-)\pi^+$ &
472 $\pm$ 24&
 42.7 $\pm$ 2.1 $\pm$ 2.2\\
\hline
\end{tabular}
\end{center}
\end{table}

Contributions to the branching fraction systematic error are shown in Table~\ref{tab:sys}. Errors due to pion tracking, particle identification and $\Ks$ reconstruction efficiency were assigned by comparing control channels in MC and data. To calculate errors due to the fit procedure, a large number of MC samples containing the amounts of signal, continuum and $B$-background events measured or fixed in data were used. The differences between the generated and fitted values using these samples were used to ascertain the sizes of any biases. Small biases of the order of a few percent were observed that were a consequence of small correlations between fit variables and were therefore assigned as systematic errors. The uncertainty of the $B$-background contribution to the fit was estimated by varying the measured branching fractions within their errors. Each background was varied individually and the effect on the fitted signal yield was added as a contribution to the uncertainty. For $B^0$ $\rightarrow$ $K^{*+}\pi^-$ there was also the $B$-background contributions from higher kaon resonances which was obtained from fits to data and added as a systematic. The uncertainty due to simulated PDFs was obtained from the channel  $B^0$ $\rightarrow$ $D^-(\rightarrow \Ks\pi^-)\pi^+$ and by varying the PDFs according to the precision of the parameters obtained from MC. In order to take correlations between parameters into account, the full correlation matrix was used when varying parameters. All PDF parameters that were originally fixed in the fit were then varied in turn and each difference from the nominal fit was combined and taken as a systematic contribution. The error in the efficiency was due to limited MC statistics, where over a million MC events were generated for the decay $B^0 \to K^0\pi^+\pi^-$ and over one hundred and fifty thousand MC events were generated for the decays $B^0$ $\rightarrow$ $K^{*+}\pi^-$ and $B^0$ $\rightarrow$ $D^-(\rightarrow \Ks\pi^-)\pi^+$. The same uncertainty due to the error in the number of \BB\ events was added to all channels. 

\begin{table}[!ht]
\caption{Summary of systematic uncertainty contributions to the branching fraction measurements $B^0 \to K^0\pi^+\pi^-$, $B^0$ $\rightarrow$ $K^{*+}\pi^-$ and $B^0$ $\rightarrow$ $D^-(\rightarrow \Ks\pi^-)\pi^+$. The errors are shown as a percentage of the measured branching fraction.}\label{tab:sys}
\begin{center}
\begin{tabular}{|c|c|c|c|}
\hline
Error &
\footnotesize{$B^0$ $\rightarrow$ $K^0\pi^+\pi^-$}&
\footnotesize{$B^0$ $\rightarrow$ $K^{*+}\pi^-$}&
\footnotesize{$B^0$ $\rightarrow$ $D^-\pi^+$}\\
source&
error(\%)&
error(\%)&
error (\%)\\
\hline
Tracking&
1.7&
1.7&
1.7\\
\hline
Particle ID &
1.9&
1.2&
3.0\\
\hline
$\Ks$ Efficiency&
4.2&
3.5&
3.0\\
\hline
Fit Bias&
4.1&
3.3&
1.8\\
\hline
$B$-background&
3.6&
9.0&
0.3\\
\hline
PDF params.&
1.5&
0.5&
0.4\\
\hline
Efficiency&
1.7&
0.6&
0.6\\
\hline

No. of \BB\ &
1.1&
1.1&
1.1\\
\hline
Total&
7.8&
10.5&
5.1\\
\hline
\end{tabular}
\end{center}
\end{table}
 Interference was also considered for the decay $B^0$ $\rightarrow$ $K^{*+}\pi^-$ where effects between the $K^{*+}(892)$ and S wave final states (non-resonant and $K^{*+}_0(1430)$) cancel and the $K^{*+}(892)$ and D wave final states ($K^{*+}_2(1430)$) cancel. This is not the case for  P wave amplitudes such as $K^{*+}_1(1410)$, yet this effect was considered negligible due to the small branching fraction of $K^{*+}_1(1410$) $\to \Ks\pi^+$ (6.6 $\pm$ 1.3\%~\cite{pdg}). 

The $CP$ violating charge asymmetry for the decay $B^0$ $\rightarrow$ $K^{*+}$$\pi^-$ was measured to be ${\cal A}_{K^*\pi} = 0.23 \pm 0.18^{+0.09}_{-0.06}$, where the first error is statistical and the second errors are systematic. The background asymmetry ${\cal A}^{Bkg}_{K^*\pi}$ was measured to be 0.01 $\pm$ 0.01 and as a further study the asymmetry ${\cal A}_{D\pi}$ for $B^0$ $\rightarrow$ $D^-(\rightarrow \Ks\pi^-)\pi^+$ was measured to be 0.00 $\pm$ 0.05 and the background asymmetry ${\cal A}^{Bkg}_{D\pi}$ was 0.06 $\pm$ 0.04, were the errors are statistical only.

The systematic error on ${\cal A}_{K^*\pi}$ was calculated by considering contributions due to track finding, particle identification, fit biases and $B$-background asymmetry uncertainties. Biases due to track finding and particle identification were found to be negligible. The fit bias contribution to the systematic error was calculated using a large number of MC samples. The contribution from $B$-background was calculated by varying the number of expected events within errors and by assuming a conservative $CP$ violating asymmetry of $\pm$ 0.5 as there are no available measurements for these decays. The resulting systematic uncertainty on the asymmetry was measured to be ${}^{+0.09}_{-0.06}$.

In summary, the branching fractions for $B^0 \to K^0\pi^+\pi^-$, $B^0$ $\rightarrow$ $K^{*+}\pi^-$ and $B^0$ $\rightarrow$ $D^-(\rightarrow \Ks\pi^-)\pi^+$ decaying to a $\Ks\pi^+\pi^-$ state have been measured and agree with previous measurements~\cite{cleo,belle,pdg}. The direct $CP$ violating parameter ${\cal A}_{K^*\pi}$ was measured for the decay $B^0$ $\rightarrow$ $K^{*+}\pi^-$ and is in agreement with the CLEO measurement~\cite{cleoacp}, with no evidence of $CP$ violation with the statistics used. Using larger datasets, one can extract amplitudes and relative phases of the resonant contributions to the Dalitz plot, with the possibility to observe new $B$-meson decays.

We are grateful for the excellent luminosity and machine conditions
provided by our \pep2\ colleagues, 
and for the substantial dedicated effort from
the computing organizations that support \babar.
The collaborating institutions wish to thank 
SLAC for its support and kind hospitality. 
This work is supported by
DOE
and NSF (USA),
NSERC (Canada),
IHEP (China),
CEA and
CNRS-IN2P3
(France),
BMBF and DFG
(Germany),
INFN (Italy),
FOM (The Netherlands),
NFR (Norway),
MIST (Russia), and
PPARC (United Kingdom). 
Individuals have received support from CONACyT (Mexico), A.~P.~Sloan Foundation, 
Research Corporation,
and Alexander von Humboldt Foundation.

\end{document}